\documentclass[sigconf]{acmart}

\AtBeginDocument{%
  }

\copyrightyear{2026}
\acmYear{2026}
\setcopyright{cc}
\setcctype{by-nc-nd}
\acmConference[SIGGRAPH Conference Papers '26]{Special Interest Group on Computer Graphics and Interactive Techniques Conference Conference Papers}{July 19--23, 2026}{Los Angeles, CA, USA}
\acmBooktitle{Special Interest Group on Computer Graphics and Interactive Techniques Conference Conference Papers (SIGGRAPH Conference Papers '26), July 19--23, 2026, Los Angeles, CA, USA}
\acmDOI{10.1145/3799902.3811214}
\acmISBN{979-8-4007-2554-8/2026/07}

\usepackage{subcaption}

\usepackage{siunitx}
\usepackage{makecell}
\usepackage{stfloats}

\usepackage{enumitem}

\begin{document}
\title{Reciprocal Latent Fields for Precomputed Sound Propagation}

\author{Hugo Seuté}
\authornote{Both authors contributed equally to this research.}
\affiliation{%
  \institution{Ubisoft La Forge}
  \city{Montréal}
  \country{Canada}
}
\email{hugo.seute@ubisoft.com}

\author{Pranai Vasudev}
\authornotemark[1]
\affiliation{%
  \institution{Audiokinetic}
  \city{Montréal}
  \country{Canada}
}
\email{pvasudev@audiokinetic.com}

\author{Etienne Richan}
\affiliation{%
  \institution{Audiokinetic}
  \city{Montréal}
  \country{Canada}
}
\email{erichan@audiokinetic.com}

\author{Louis-Xavier Buffoni}
\affiliation{%
  \institution{Audiokinetic}
  \city{Montréal}
  \country{Canada}
}
\email{xbuffoni@audiokinetic.com}

\renewcommand{\shortauthors}{Seuté et al.}

\begin{abstract}

Realistic sound propagation is essential for immersion in a virtual scene, yet physically accurate wave-based simulations remain computationally prohibitive for real-time applications. Wave coding methods address this limitation by precomputing and compressing impulse responses of a given scene into a set of scalar acoustic parameters, which can reach unmanageable sizes in large environments with many source-receiver pairs. We introduce Reciprocal Latent Fields (RLF), a memory-efficient framework for encoding and predicting these acoustic parameters.
The RLF framework employs a volumetric grid of trainable latent embeddings decoded with a symmetric function, ensuring acoustic reciprocity. We study a variety of decoders and show that leveraging Riemannian metric learning leads to a better reproduction of acoustic phenomena in complex scenes. Experimental validation demonstrates that RLF maintains replication quality while reducing the memory footprint by several orders of magnitude. Furthermore, a MUSHRA-like subjective listening test indicates that sound rendered via RLF is perceptually indistinguishable from ground-truth simulations.
\end{abstract}


\begin{teaserfigure}
  \includegraphics[width=\textwidth]{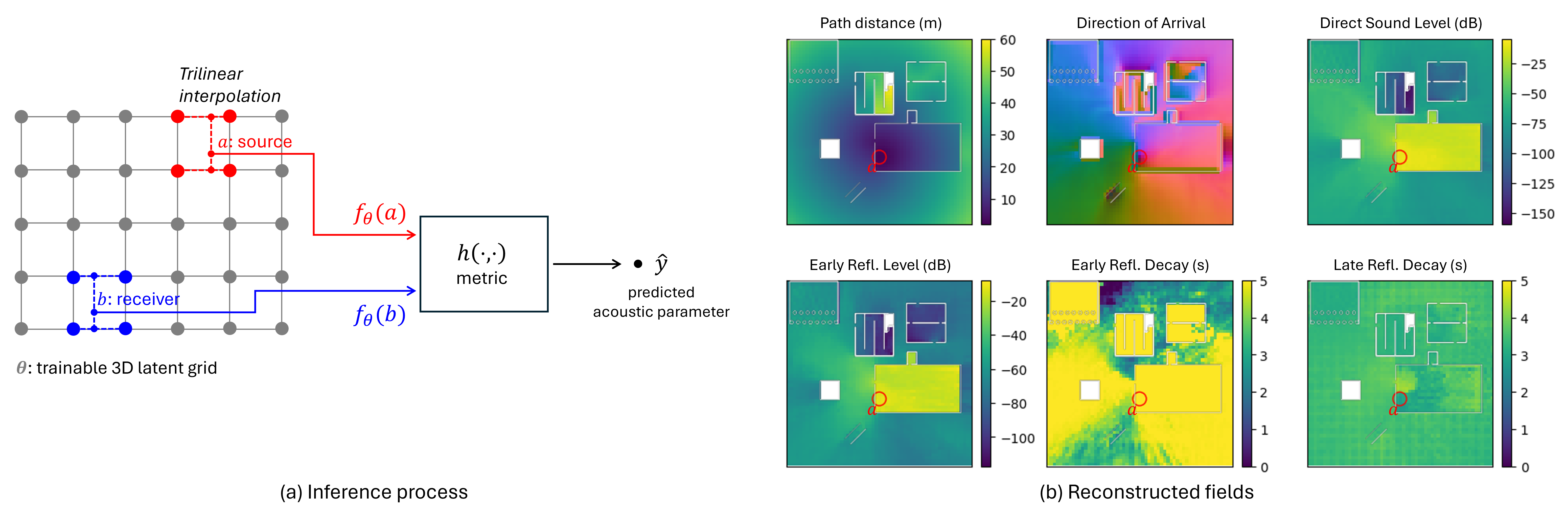}
  \caption{Diagram of a Reciprocal Latent Field model and its predicted acoustic parameters for a fixed source position in the \emph{Audio Gym} map.}
  \Description{Model diagram for inference and reconstructed fields for a Reciprocal latent field model}
  \label{fig:teaser}
\end{teaserfigure}


\maketitle

\section{Introduction} \label{sec:introduction}
In interactive virtual environments, such as video games, audio is a primary driver of immersion and a critical tool for directing attention within a 3D space. Realistic auditory experiences require accurate modeling of complex wave phenomena, specifically diffraction around obstacles and reverberation. These effects are physically characterized by the impulse response (IR), which captures the spectral and temporal transformations a unit impulse undergoes as it propagates from source to receiver.

Because the IR changes with every movement of the source or receiver, accurately simulating these wave phenomena in real-time is beyond the computational budgets of game engines. As detailed in Sec. \ref{sec:related_work}, geometric methods such as ray-tracing remain CPU-intensive for complex scenes, while rooms and portals systems require labor-intensive manual markup. Our work builds upon the wave coding method \cite{Raghuvanshi2014}, which addresses these issues via precomputation of acoustic parameters (detailed in Sec. \ref{sec:parametric_coding}). However, this approach scales poorly to large environments, as it requires computing and storing data for all necessary source-receiver configurations.

To address this bottleneck, we introduce \emph{Reciprocal Latent Fields (RLF)} in Sec. \ref{sec:reciprocal_lf}. While applied here to wave coding, RLF is a generalized framework that compresses reciprocal scalar fields into a grid of $n$-dimensional embeddings forming a latent manifold. We employ decoders to predict field values as geodesic distances between arbitrary source and receiver embeddings, similar to \cite{Zhang2023}. By restricting these decoders to symmetric functions, our method guarantees acoustic reciprocity by design.

We summarize our main contributions as follows:
\begin{itemize}[leftmargin=1.0em, style=nextline, noitemsep]
    \item \emph{Reciprocal Latent Fields (RLF)}. A novel method for encoding acoustic paths as metrics over a latent manifold, inherently enforcing physical reciprocity (Sec. \ref{sec:reciprocal_lf:idea}).
    \item \emph{Riemannian Decoder Architecture}. A decoder that significantly improves reconstruction accuracy over simpler baselines with negligible computational overhead by including a local metric tensor to warp space (Sec. \ref{sec:reciprocal_lf:idea:riemannian}).
    \item \emph{Lightweight decoders for acoustic parameters}. An extension of RLF to a complete set of acoustic parameters, including non-metric quantities such as energy levels and decay times, enabling real-time, memory-efficient acoustic rendering (Sec. \ref{sec:reciprocal_lf:sound_propagation}).
    \item \emph{Benchmark and study}.  In Sec. \ref{sec:results}, we compare decoder designs and embedding dimensionality. In conjunction with a subjective study, our results demonstrate that the Riemannian RLF approach maintains high fidelity while compressing wave coding data by several orders of magnitude.
\end{itemize}

\section{Related Work} \label{sec:related_work}
\subsection{Impulse response generation} \label{sec:related_work:ir_gen}
Existing methods for computing acoustics generally focus on predicting full IRs, either via physical simulation or machine learning approximation.

\subsubsection{Traditional methods} \label{sec:related_work:ir_gen:traditional}
Physical approaches are split into geometrical and wave-based methods. Geometrical methods, such as image source \cite{Kuttruff2000} and acoustic radiance transfer \cite{Siltanen2010, Nosal2004} are computationally tractable but require costly approximations to model diffraction \cite{Tsingos2001, Savioja2015}. Conversely, wave-based solvers (e.g., FDTD \cite{Hamilton2021}, FEM \cite{Melander2024}, BEM \cite{Hargreaves2019}) inherently capture wave phenomena but remain too computationally intensive for real-time applications.

\subsubsection{Machine learning-based methods} \label{sec:related_work:ir_gen:ml}
Recent work accelerates IR generation using mesh-based encoders \cite{Ratnarajah2022a, Li2025}, neural operators \cite{Lu2021, BorrelJensen2024}, or implicit neural representations \cite{Luo2022}. However, predicting raw IRs remains ill-suited for video games. First, massive parameter counts lead to prohibitive inference costs \cite{BorrelJensen2024, Kelley2024}. Second, raw IRs lack spatial smoothness, resulting in interpolation artifacts and wall leakage. Third, they lack the tunable parameters (e.g., loudness, decay) required by sound designers. Finally, generalization is often limited by training on residential datasets \cite{Chen2020, Tang2022} or simplified geometries \cite{BorrelJensen2024, Karakonstantis2023}, restricting applicability in complex game environments.

\subsection{Simulating sound propagation in games} \label{sec:related_work:simulation}
The common systems for sound propagation in games are ray-tracing, rooms and portals, and wave coding.

\subsubsection{Ray-tracing} \label{sec:related_work:simulation:ray_tracing}
Ray-tracing models propagation via specular reflections at boundary surfaces ignoring phase coherence and fails to inherently model diffraction \cite{Raghuvanshi2023}. Real-time performance is bottlenecked by the visibility tree, which defines the valid reflection paths between a source and receiver and requires expensive re-calculations whenever the source or receiver moves \cite{Bertram2005, Taylor2012}. While optimizations like diffuse rain exist \cite{Schroeder2011, Pelzer2011, Thomas2017}, they do not resolve this computational bottleneck. Furthermore, valid reflection path finding is unstable in complex geometries, causing audio dropouts when ray density is insufficient \cite{Taylor2012, Savioja2015}.

\subsubsection{Rooms and portals systems} \label{sec:related_work:simulation:rooms}
Stemming from classical work on acoustically coupled rooms \cite{Kuttruff2000}, rooms and portals systems decompose the virtual scene into discrete volumes (rooms) connected by openings (portals), allowing for simplified propagation modeling. \cite{Stavrakis2008} proposes a topological approach using reverberation graphs to model energy exchange between coupled volumes, achieving update rates (10--100 Hz) suitable for real-time use. A prominent commercial implementation is Wwise Rooms and Portals \cite{Audiokinetic2026}. While computationally efficient, this approach is labor-intensive, requiring manual markup of room boundaries and reverberation parameters. Moreover, this abstraction breaks down in outdoor or hybrid environments where acoustic boundaries are ambiguous \cite{Raghuvanshi2023}.

\subsubsection{Wave coding systems} \label{sec:related_work:simulation:wave_coding}

Wave-coding \cite{Raghuvanshi2014} shifts the computational burden of wave-based sound propagation to an offline simulation. Simulations are run using a representative set of emitters and a dense grid of receiver probes. Compact acoustic parameters are then extracted from the IRs computed at each probe to drive runtime filters (loudness, reflection energy, decay times, direction). This method inherently accounts for diffraction without manual room definition, offering a physically accurate yet automated alternative to rooms and portals systems. The main limitation of this approach is the memory footprint which scales proportionally to the product of the number of emitters and receivers. Even with compression, the memory footprint scales poorly for large maps on memory-constrained devices. We address this memory bottleneck by replacing explicit parameter grids with our Reciprocal Latent Field framework.

\subsection{Distance encoding methods}\label{sec:related_work:distance_encoding}
Our method for sound propagation builds on the idea of distance encoding. Many solutions exist to compute geodesics, spanning a range of memory-compute trade-offs. At one end, precomputed distance matrices provide \(O(1)\) query time at a cost of \(O(N^2)\) memory, where \(N\) is the number of query locations. At the other end, runtime methods such as Fast Marching \cite{Sethian1999} require little memory but incur \(O(N \log N)\) computation. Methods such as distance oracles \cite{Thorup2005} and the heat method \cite{Crane2013} provide intermediate trade-offs. Another approach is distance embedding, such as \cite{Zhang2023}, which learns an \(O(N)\)-memory representation and decodes it in \(O(1)\) time, while introducing some error. These properties are attractive for real-time acoustics, where memory and compute budgets are limited and approximations can be tolerated. Our RLF method falls in this distance embedding approach using a lightweight metric-based decoder that offers accurate reconstruction and preserves spatial gradients.

\section{Parametric wave field coding for precomputed sound propagation} \label{sec:parametric_coding}
As there is no publicly available implementation of the wave coding pipeline \cite{Raghuvanshi2014}, we describe our own.

\subsection{Acoustic Simulation} \label{sec:parametric_coding:simulation}
We perform wave-based acoustic simulations using the \emph{Pretty Fast FDTD} (PFFDTD) solver \cite{Hamilton2021}. Given the scene mesh, PFFDTD voxelizes the domain into a regular grid with spacing
$h = c / (f_{\max} \cdot \mathrm{PPW})$, where $c$ is the speed of sound, $f_{\max}$ the maximum simulated frequency, and PPW the number of points per wavelength.  
The simulation time step is $\Delta t = S h / c$, where $S$ is the Courant factor \cite{Courant1967}.
Each voxel is assigned the acoustic impedance of the nearest intersecting triangle.  
Material properties are specified as Sabine absorption coefficients over 11 octave bands spanning 16~Hz to 16~kHz.

For each simulation, a unit impulse is emitted at a source position and the discretized acoustic wave equation is solved at every grid point over time. The output is a 3D array of impulse responses sampled at receiver locations arranged in a regular grid.
In all experiments, we use $c = 343$~m/s, $f_{\max} = 500$~Hz, $\mathrm{PPW} = 5$, and $S = 1/\sqrt{3}$, yielding a spatial resolution of 0.137~m and a time step of $2.3 \times 10^{-4}$~s.  
Impulse responses are simulated for a duration of two seconds.

\subsection{Parameter Estimation} \label{sec:parametric_coding:parameter_estimation}
For a given source position $a$, we extract a set of acoustic parameters for each receiver position $b$: the wave path distance $\pi(a,b)$, the direct sound level $L_{\text{DS}}(a,b)$, the early reflections level $L_{\text{ER}}(a,b)$, the early and late reflection decay times $\tau_{\text{ER}}(a,b)$ and $\tau_{\text{LR}}(a,b)$, and the direction of arrival vector $\delta(a,b)$.

As in \cite{Raghuvanshi2014}, levels and decay times are derived from the energy density of the band-pass filtered IR. Decay times are estimated from the Schr\"oder integral $S(t)$ measuring the energy remaining in the IR from time $t$ onwards \cite{Schroeder1965}.

\subsubsection{Path distance} \label{sec:parametric_coding:parameter_estimation:path_distance}
The distance traveled by the wave from source to receiver is estimated by $\pi = c (t_{DS} - t_0)$,
where $t_0$ is the emission time and the arrival time $t_{\text{DS}}$ is detected using a relative threshold on the unfiltered IR. Due to diffraction, $\pi(a,b)$ represents the shortest path between $a$ and $b$, accounting for detours around obstacles. While the path distance isn't used directly for sound rendering, it is used to estimate the direction of arrival. 

\subsubsection{Direction of arrival}  \label{sec:parametric_coding:parameter_estimation:doa}
The direction of arrival (DOA) is estimated from the spatial gradient of the path distance field with respect to the receiver position $b$:
\begin{equation} \label{eq:doa_definition}
    \delta(a, b) = -\nabla_{b} \pi(a, b) / \big\|\nabla_{b} \pi(a, b)\big\|.
\end{equation}
Because receivers lie on a regular grid, it is straightforward to approximate $\nabla_b \pi$ using finite differences.

\subsubsection{Direct sound level}  \label{sec:parametric_coding:parameter_estimation:ds_level}
The direct sound level measures the loudness of the sound that first arrives at the listener. 
This parameter captures distance attenuation and obstruction effects. $L_{\text{DS}}$ is formally defined as the energy density integrated over the direct sound window
$[t_{\text{DS}},\, t_{\text{DS}} + 15\,\mathrm{ms}]$, expressed in decibels.

\subsubsection{Reflection levels} \label{sec:parametric_coding:parameter_estimation:er_level} The early reflection level measures the loudness of specular reflections arriving right after the direct sound.
$L_{\text{ER}}$ is computed analogously to $L_{\text{DS}}$, but over the early reflections window $[t_{\text{DS}} + 15\mathrm{ms} , t_{\text{DS}} + 115\mathrm{ms}]$.
The late reflection level $L_{\text{LR}}$ measures the loudness of the reverberation tail. $L_{\text{LR}}$ is calculated such that the energy in the final 25~ms of the ER window matches the energy in the first 25~ms of the LR window $[t_{\text{DS}} + 415\mathrm{ms} , t_{\text{DS}} + 1015\mathrm{ms}]$, ensuring a smooth transition between regimes.
Together with decay times, these levels capture properties of the acoustic environment related to enclosure, room size, and surface absorption or reflectivity.

\subsubsection{Reflection decay times}
The decay times $\tau_{\text{ER/LR}}$ give a measure of how long it takes for the early and late reflections to fade out. $\tau_{\text{ER}}$ and $\tau_{\text{LR}}$ are estimated from the slope $s$ of the Schr\"oder integral $S(t)$ over their respective temporal windows. The decay slope is converted to a decay time via
$\tau_{\mathrm{ER/LR}}=-60/s_{\mathrm{ER/LR}}$.
For early reflections, the slope $s_{\text{ER}}$ is estimated as the root mean square of forward differences of $S(t)$ over the ER window. Unlike \cite{Raghuvanshi2014}, which define decay bounds using relative energy thresholds, we estimate $\tau_{\text{ER}}$ over the same fixed time interval used to compute $L_{\text{ER}}$. For late reflections, the slope $s_{\text{LR}}$ is obtained by linear regression of $S(t)$ over the LR window.

\subsection{Rendering} \label{sec:parametric_coding:rendering}
A sound source is rendered by summing three signal paths: the unaffected direct sound component (or \emph{dry} path) and two reverberant (or \emph{wet} path) components, corresponding to the early and late reflections.

\subsubsection{Reference IR selection} \label{sec:parametric_coding:rendering:reference_IR}
We choose six reference IRs for convolutional filters $P^{\text{\{ER, LR\}}}_{\{S, M, L\}}$, which model the early and late reflections for small, medium, and large spaces.
Unlike \cite{Raghuvanshi2014}, who synthesize canonical IRs, we select real recorded impulse responses characterized by decay times $\tau^{\text{\{ER, LR\}}}_{\{S, M, L\}}$. This provides an extra degree of control for sound designers who may choose different IRs based on perceptual and aesthetic criteria. 

\subsubsection{Dry path} \label{sec:parametric_coding:rendering:dry_path}
The dry path signal is obtained by scaling the input signal $x_{\text{in}}(t)$ by the direct sound level:
\begin{equation} \label{eq:dry_path_rendering}
x_{\text{dry}}(t) = 10^{L_{\text{DS}}/20} \, x_{\text{in}}(t).
\end{equation}

\subsubsection{Wet paths} \label{sec:parametric_coding:rendering:wet_path}
The early reflections component is rendered as
\begin{equation} \label{eq:wet_path_rendering}
x_{\text{wet}}^{\text{ER}}(t) =
10^{L_{\text{ER}}/20}
\sum_{j \in \{\text{S},\text{M},\text{L}\}}
\left( x_{\text{in}}(t) \ast P^{\text{ER}}_j(t) \right) \sqrt{w_{\text{ER},j}},
\end{equation}
 where weights $w_{\text{ER},j}$ are obtained by piecewise linear interpolation as a function of the estimated decay time $\tau_{\text{ER}}$. For example, if $\tau_{\text{ER}}$ = 2.0 s and $\tau^{\text{ER}}_{\{S, M, L\}} = \left\{ 0.5, 1.0, 3.0 \right\}$ s, then $w_{\text{ER},\{S, M, L\}} = \left\{0.0, 0.5, 0.5 \right\}$. Late reflections are rendered analogously using $\tau_{\text{LR}}$, $L_{\text{LR}}$ and $P^{\text{LR}}_{\{S, M, L\}}$.

\subsubsection{Runtime interpolation of parameters} \label{sec:parametric_coding:rendering:interpolation}
The acoustic parameters described in Sec.~\ref{sec:parametric_coding:parameter_estimation} are stored on a discrete grid. At runtime, parameters are obtained for a given position $p$ via linear interpolation over neighboring grid points. To prevent interpolation artifacts across obstacles, grid points that do not have direct line of sight to $p$ are excluded.

\subsubsection{Spatialization} \label{sec:parametric_coding:rendering:spatialization}
Dry signals are spatialized using Vector Base Amplitude Panning (VBAP) \cite{Pulkki1997}, driven by the direction of arrival vector $\delta$. 
Reverberant signals are spatialized based on the assumption that reverberant energy arrives from many directions but remains statistically biased toward the direct path. One third of the wet signal energy is spatialized using the same direction as the dry path, while the remaining energy is rendered omnidirectionally.

\section{The Reciprocal Latent Field Framework} \label{sec:reciprocal_lf}
\subsection{Reciprocal Latent Fields} \label{sec:reciprocal_lf:idea}

We focus first on efficiently estimating the \emph{path distance} $\pi(a,b)$ for arbitrary source–receiver pairs $(a,b)$. Extensions to other acoustic parameters are detailed in Sec.~\ref{sec:reciprocal_lf:sound_propagation}.

While \cite{Raghuvanshi2014} encode these distances as 6D fields (three dimensions for each of source $a$ and receiver $b$), which are sampled by 6D linear interpolation, we instead propose to model the problem with two coupled 3D mappings: 
\begin{equation} \label{eq:latent_recomposition}
    \hat{\pi}(a, b) = h\big(f(a), \; f(b)\big),
    \quad f: \mathbb{R}^3 \to \mathbb{R}^n
\end{equation}

Here, $f$ maps 3D positions to $n$-dimensional latent embeddings, and $h$ is a metric in this latent space, ensuring acoustic reciprocity.
If $f$ and $h$ have been properly determined, Eq. \eqref{eq:latent_recomposition} yields direct estimates of the path distance at runtime without requiring explicit path-finding or simulation for every possible source–receiver pair.

We implement $f$ as a trilinear interpolation over a grid of trainable vectors $\theta\in \mathbb{R}^{H \times W \times D \times n}$ :
\begin{equation} \label{eq:latent_grid_interpolation}
    f_\theta(a) = \mathrm{interp}(\theta, a_x, a_y, a_z),
\end{equation}
To prevent any leakage across walls, the interpolation is restricted to vertices that are visible from the query point, as described in Sec.~\ref{sec:parametric_coding:rendering:interpolation}. Unlike implicit neural representations \cite{Sitzmann2020, Mueller2022}, which tend to smooth high frequencies \cite{Fan2024}, this grid-based approach preserves the sharp discontinuities critical for modeling diffraction around obstacles. We refer to this formulation as \emph{Reciprocal Latent Fields (RLF)}:
\begin{equation} \label{eq:rlf_definition}
    \hat{\pi}(a, b) = h( f_\theta(a), \;  f_\theta(b) ).
\end{equation}
Model parameters $\theta$ are optimized to approximate the simulated ground-truth distances $\pi(a, b)$ via gradient descent. The loss is the mean squared error over sets of source ($\mathcal{A}$) and receiver ($\mathcal{B}$) positions:
\begin{equation} \label{eq:training_loss}
    \mathcal{L}(\theta) =
    \frac{1}{|\mathcal{A}||\mathcal{B}|}
    \sum_{a \in \mathcal{A}} \sum_{b \in \mathcal{B}}
    \big( \hat{\pi}(a, b) - \pi(a, b) \big)^2.
\end{equation}

We remark that if $h$ is a pseudo-metric, then $\hat{\pi}$ also has the properties of a pseudo-metric: non-negativity, identity, symmetry and triangle inequality \cite{Gruffaz2025}. 

\subsubsection{Euclidean RLF}  \label{sec:reciprocal_lf:idea:euclidean}
The simplest realization of $h$ is the Euclidean distance:
\begin{equation}
    \label{eq:euclidean_model}
    \hat{\pi}_{EUC}(a, b) = \big\| f_\theta(a) \;-\; f_\theta(b) \big\|.
\end{equation}

\begin{figure}[htbp]
  \centering
  \includegraphics[width=\linewidth]{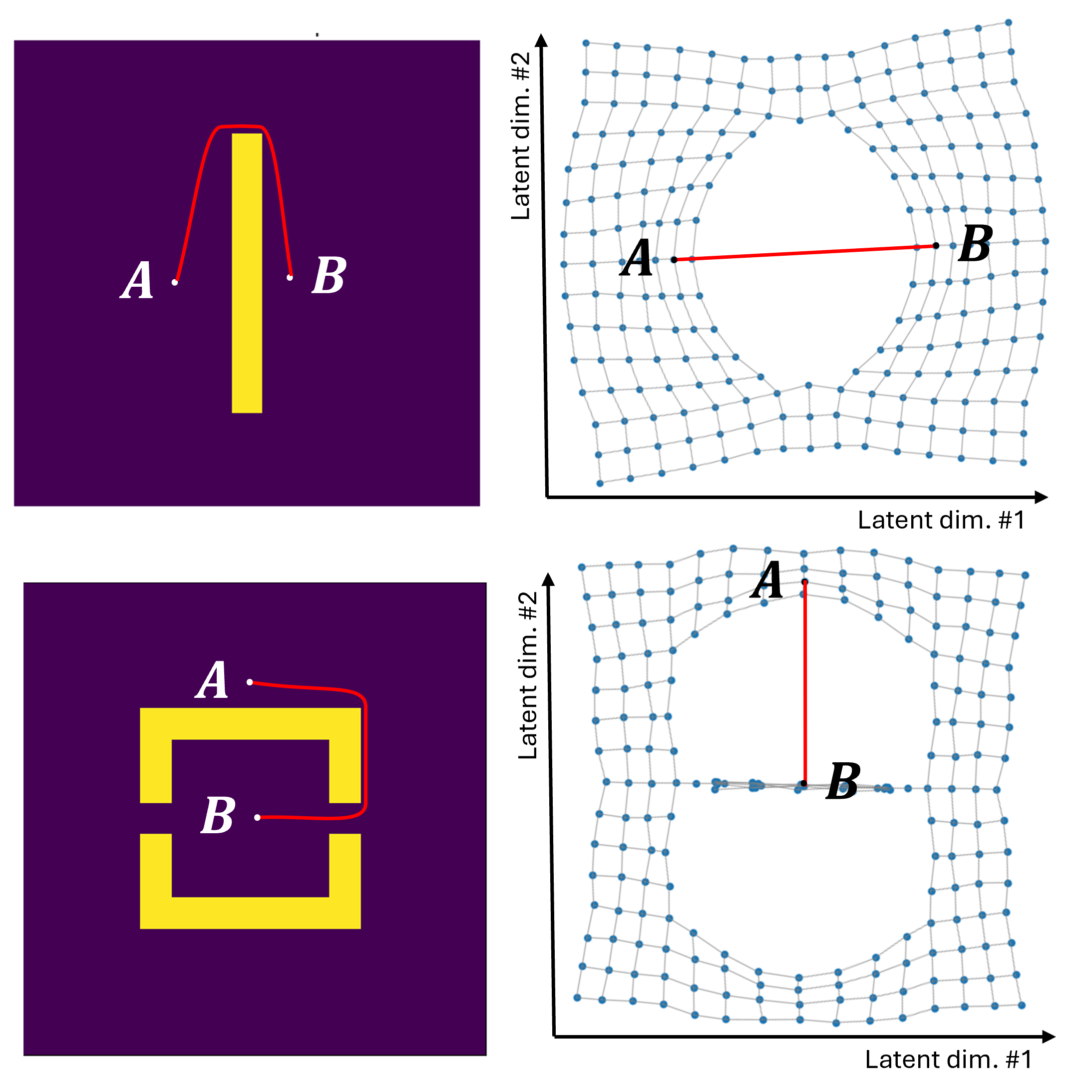}
  \caption{Visualization of the 2D latent space from an Euclidean RLF trained to reproduce path distance, on two different 2D geometries}
   \label{fig:2D_RLF}
  \Description{Visualization of the 2D latent space from an Euclidean RLF trained to reproduce path distance, on two different 2D geometries}
\end{figure}

Fig.~\ref{fig:2D_RLF} illustrates 2D latent spaces trained on two toy problems. The grid's regularity indicates that the Euclidean RLF approximates distances well in open areas (e.g. in the corners of the grid). However, the latent embeddings become over-constrained near obstacles. This results in visible warping, as the 2D space cannot reconcile conflicting path length constraints. We observe that increasing the number of latent dimensions $n$ reduces errors to some extent, but this effect saturates quickly (see Sec.~\ref{sec:results:latent_space_ablation}).

\subsubsection{Riemannian RLF}  \label{sec:reciprocal_lf:idea:riemannian}

With a fixed number of latent dimensions in the Euclidean RLF, the large number of pairwise constraints can become mutually inconsistent, creating an overconstrained optimization problem.
To address this, we extend the formulation to the framework of \emph{Riemannian metric learning} \cite{Gruffaz2025}, allowing a local metric $G$ to vary across the latent space. 
This relaxes the global constraints of the Euclidean formulation, enabling the model to locally expand or contract the latent manifold to satisfy the training relations.

In latent space, the Riemannian metric between points $a$ and $b$ is defined as
\begin{equation} \label{eq:full_riemannian}
    d_G(a,b) = \inf_{\gamma:[0,1]\to \mathbb{R}^n} \int_0^1 \sqrt{ \dot{\gamma}(t)^\top G(\gamma(t)) \dot{\gamma}(t) } \, dt,
\end{equation}
where $\gamma[0,1]$ is a smooth curve in the latent space between $f_\theta(a)$ and $f_\theta(b)$, $\dot{\gamma}(t)$ is the tangent vector to the curve at $t$, and $G(\gamma(t))$ is a symmetric positive definite metric tensor representing the local coordinate system. The infimum selects the shortest possible curve $\gamma$, also called a geodesic.

We model the path distance by a Riemannian metric in the latent space, where the matrix field $G$ is learned jointly with our latent grid $\theta$:
\begin{equation} \label{eq:riemann_model}
    \hat\pi_{RIE}(a, b) = d_G(f_\theta(a), f_\theta(b)).
\end{equation}

Since solving Eq.~\eqref{eq:full_riemannian} directly is complex and requires an optimization process \cite{Gruffaz2025}, we employ a first-order linearization at the midpoint $m = (f_\theta(a)+f_\theta(b))/2$, yielding an approximation without the need for integration and infimum search:
\begin{equation} \label{eq:riemann_midpoint}
    d_G(a, b) \approx \sqrt{(f_\theta(a) - f_\theta(b))^\top G(m) (f_\theta(a) - f_\theta(b))}.
\end{equation}
We observe empirically that it remains accurate even for long-range distances (see Sec.~\ref{sec:results:comparative}), likely due to the joint optimization of the latent space with the metric $G$. Eq. \eqref{eq:riemann_midpoint} can be interpreted as a Mahalanobis distance with a local covariance matrix, which reverts to Euclidean distance Eq.~\eqref{eq:euclidean_model} when $G$ is the identity matrix.

In practice, we model $G(m)$ using a simple trainable linear layer $ \Lambda$ without bias. 
We propose two variants: a full positive semi-definite model
\begin{equation}
 G_{PSD}(m) = \Lambda(m)^\top \Lambda(m),
\quad \Lambda: \mathbb{R}^n \to \mathbb{R}^{n \times n},
\end{equation}
and a lighter diagonal alternative 
\begin{equation}
    G_{DIAG}(m) = \mathrm{diag}\big(\lambda(m)\big)^2,
    \quad \lambda: \mathbb{R}^n \to \mathbb{R}^{n} ,
\end{equation}
where $\lambda$ is an $n^2$-parameter linear layer. In this case, the distance in Eq.~\eqref{eq:riemann_midpoint} can be further reduced to a locally-weighted Euclidean distance.

\begin{equation}
\label{eq:riemann_diag}
    d_{G_{DIAG}}(a, b) = \sqrt{ \sum_{i=1}^n \lambda(m)_i^2 \, (f_\theta(a)_i - f_\theta(b)_i)^2 }.
\end{equation}

We compare the diagonal ($G_{DIAG}$) and full PSD ($G_{PSD}$) models in Sec.~\ref{sec:results:latent_space_ablation}, showing that the full model offers a more precise reconstruction, while the diagonal model still offers a good approximation at a fraction of the computational cost.

\subsubsection{Distance field with MLPs} \label{sec:reciprocal_lf:idea:mlp}

Alternatively, we can model the decoder function directly using a multilayer perceptron (MLP) $\phi$, which maps concatenated latent pairs to a scalar. To ensure acoustic reciprocity, we symmetrize the input:
\begin{equation} \label{eq:mlp_model}
  \hat{\pi}_{MLP}(a,b) = \frac{1}{2} \Big( \phi\big([f_\theta(a) \, \| \, f_\theta(b)]\big) + \phi\big([f_\theta(b) \, \| \, f_\theta(a)]\big) \Big), 
\end{equation}
where $[ \cdot \|\, \cdot ]$ denotes the concatenation of the latent vectors and $\phi$ is the MLP function.

The advantage of using an MLP is its flexibility: it can, in principle, learn complex nonlinear interactions between latent variables. However, the lack of built-in metric properties may degrade generalization compared to metric-based approaches (Sec.~\ref{sec:results:comparative}).

\subsection{Sound propagation specific fields} \label{sec:reciprocal_lf:sound_propagation}
We now extend the RLF framework to the remaining acoustic parameters described in Sec.~\ref{sec:parametric_coding:parameter_estimation}. Unlike the path distance $\pi(a,b)$, other parameters do not inherently behave as metrics and require specific adaptations to fit the RLF framework. Unless otherwise noted, each parameter utilizes its own distinct grid of trainable latent vectors $\theta$.

\subsubsection{Sound Levels} \label{sec:reciprocal_lf:sound_propagation:levels}
In free-field conditions, the sound pressure level decreases with distance according to the inverse-square law $L_0-20\log_{10}\!\big(\|a - b\|\big)$, where $L_0$ denotes the reference level at a unit distance.
We replace the attenuation term with a learned latent metric that captures complex propagation effects. Quantities are optimized directly in decibels, removing the need for explicit logarithmic terms in the model. We observed that learning the attenuation term in log-scale yields a more consistent formulation and improved stability in our experiments.
We model the \emph{Direct Sound (DS)} as a global reference level $L_0$ attenuated by the distance in the embedding space:
\begin{equation} \label{eq:ds_level_model}
    \hat{L}_{\mathrm{DS}}(a,b) = L_0 - h(f_\theta(a),f_\theta(b)).
\end{equation}

For \emph{Early Reflections (ER)}, the strength of reflections depends on the local geometry (e.g., reverberant vs. open spaces). To capture this spatial variability, we replace the constant $L_0$ with a position-dependent local level field $L_0(x)$:
\begin{equation} \label{eq:er_lr_level_model}
    \hat{L}_{\mathrm{ER}}(a,b) = \frac{L_0(a) + L_0(b)}{2} - h(f_\theta(a),f_\theta(b)).
\end{equation}
Averaging $L_0$ at the source and receiver preserves reciprocity. We model the local level field as a linear projection of the latent embedding $L_0(x) = w^\top f_\theta(x) + \beta$, where $w$ and $\beta$ are learned jointly with the grid.

\subsubsection{Decay Times} \label{sec:reciprocal_lf:sound_propagation:decay}

Decay times cannot be naturally expressed as distances, but still require reciprocity. We model them through a dot product between the latent embeddings of the receiver and source:
\begin{equation} \label{eq:dot_product_decay_times}
    \hat{\tau}_{\mathrm{ER/LR}}(a,b)
    = K \sigma\!\big(f_\theta(a)^\top f_\theta(b)\big),
\end{equation}
where $\sigma$ is the sigmoid function and $K$ is a constant that defines the longest decay time permitted by the rendering scheme (see Sec.~\ref{sec:parametric_coding:rendering}). We did not pursue the Riemannian generalization of the dot product $f_\theta(a)^\top G(m) f_\theta(b)$ as initial experiments showed no practical improvement over this simpler formulation.

\subsubsection{Direction of Arrival} \label{sec:reciprocal_lf:sound_propagation:doa}

Unlike the other acoustic parameters, the DOA is \emph{not} reciprocal: in general, $\delta(a,b) \neq -\delta(b,a)$. Hence it cannot be obtained as the direct output of an RLF model. As in Sec.~\ref{sec:parametric_coding:parameter_estimation:doa}, we instead estimate it from the gradient of the learned path distance field $\hat{\pi}$ (see Eq. \eqref{eq:doa_definition}). This field must be accurate and sufficiently smooth, as noise or discontinuities will lead to unstable DOA predictions

\subsection{Practical setup}  \label{sec:reciprocal_lf:practical}

\subsubsection{Shared latent space} \label{sec:reciprocal_lf:practical:shared_latent}

To balance memory efficiency with reconstruction accuracy, we group acoustic parameters into three families, each assigned a dedicated latent grid: (1) Path distance $\theta_\pi$, (2) Levels $\theta_L$ ($L_{\mathrm{DS}}, L_{\mathrm{ER}}$), and (3) Decay times $\theta_\tau$ ($\tau_{\mathrm{ER}}, \tau_{\mathrm{LR}}$). This separation allows us to tune the spatial resolution of each parameter group independently.

Latent spaces of multi-parameter groups  (Levels and Decay) are mapped to $k=2$ distinct acoustic parameters.
When using the parameter-free \emph{Euclidean RLF}, we add a trainable linear projection $\mathbb{R}^n \to \mathbb{R}^{k \times n}$ to the decoder. In contrast, \emph{Riemannian} and \emph{MLP} models handle this adaptation via their trainable decoder weights.

\subsubsection{Dataset Generation} \label{sec:reciprocal_lf:practical:dataset}

Training data is generated by the simulation pipeline described in Sec.~\ref{sec:parametric_coding:simulation}.
For each source position, this produces five volumetric acoustic parameter grids ($\pi$, $L_{DS}$, $L_{ER}$, $\tau_{ER}$, and $\tau_{LR}$) covering all receiver locations.

We employ an adaptive source sampling algorithm to maximize spatial coverage while minimizing redundant full-wave simulations. We wish to densely sample acoustically complex areas (e.g., a maze) and sparsely sample open regions (e.g., a football field). Our algorithm uses a 3D binary grid to represent scene geometry, where voxels are marked as either “free” or “obstacle". First, the set of selected sources is initialized by randomly choosing 20 positions from the free voxels. Then, all voxels that are directly visible from any of these sources are marked as “obstacle.” Sources are subsequently added one by one by randomly sampling the remaining voxels that are not visible from any previously selected source. This process repeats until no free voxels remain, ensuring that the selected source set provides full line-of-sight coverage of the scene. The algorithm is stochastic and can be run multiple times to obtain different source sets.

\begin{figure}[htbp] 
  \centering
  \includegraphics[width=\linewidth]{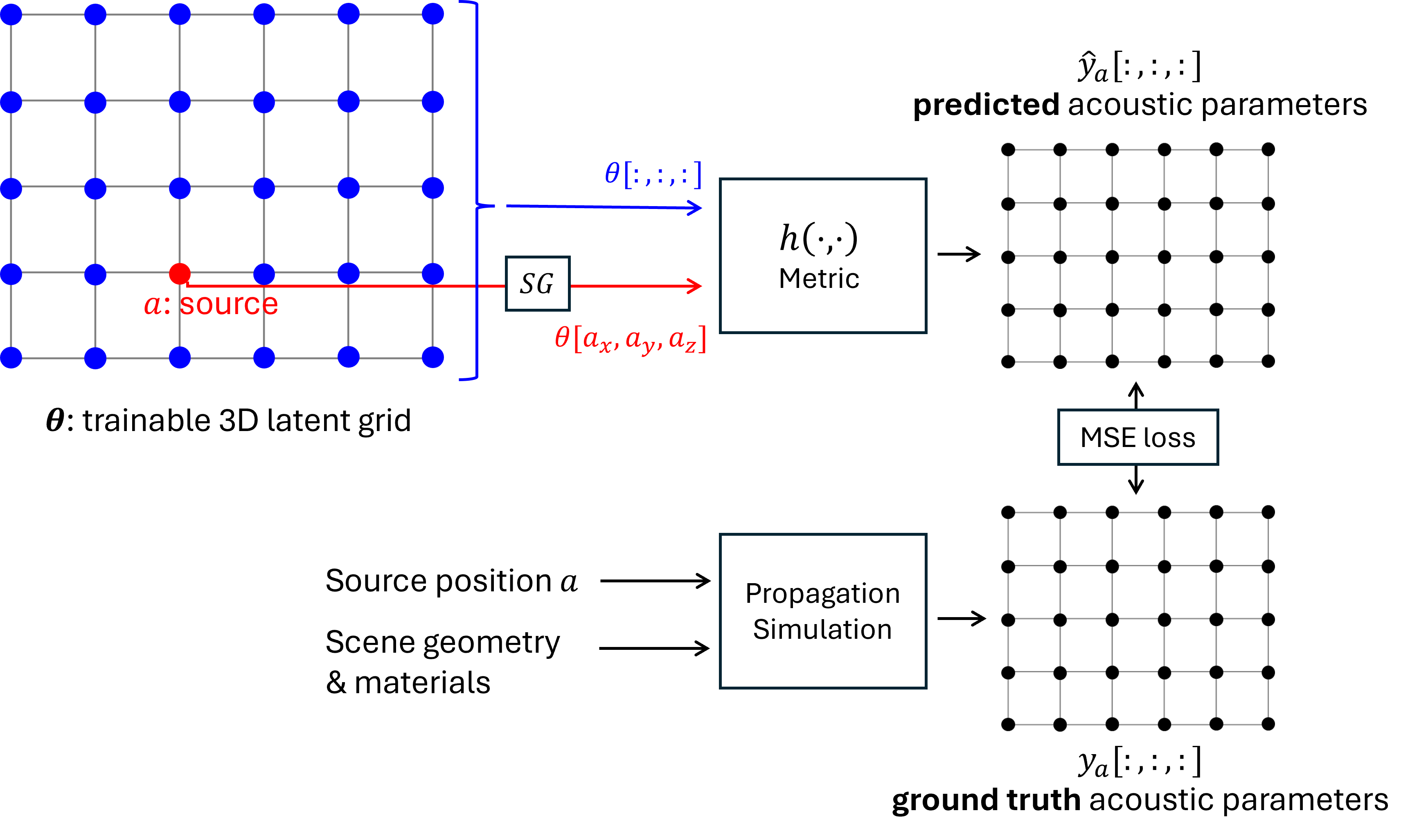}
  \caption{Diagram of a RLF model during the training phase. \textnormal{\emph{SG} designates the stop gradient operator.}}
  \label{fig:system_diagram}
  \Description{Diagram of the proposed system}
\end{figure}

\subsubsection{Training}  \label{sec:reciprocal_lf:practical:training}

We minimize the Mean Squared Error (MSE) between the predicted fields and the simulated ground truth (see Fig. \ref{fig:system_diagram}). The three parameter groups $\theta_\pi$, $\theta_L$ and $\theta_\tau$ can be trained jointly or independently. During training, a batch containing data for $N_B$ sources is assembled, and the model predicts the corresponding volumetric fields. This enables more efficient data access compared to individually sampling source-receiver pairs. Because the decoder $h$ is symmetric and shares the same mapping $f_\theta$ for source and receiver, the model naturally generalizes to source positions unseen during training.

In practice, we found that two adjustments improve convergence and reduce artifacts in the reconstruction.  
First, we use a learning rate for the latent grids $\theta$ that is an order of magnitude smaller than that of the decoder $h$, which allows the metric to adapt to a more stable geometric structure.  
Second, we stop gradients from back-propagating through the latent vector located at the source position (see Fig. \ref{fig:system_diagram}), preventing its overfitting.

\section{Results}  \label{sec:results}
\subsection{Experimental Setup}  \label{sec:results:setup}

\subsubsection{Maps}  \label{sec:results:setup:maps}
\begin{figure*}[]
  \centering
  \includegraphics[width=\linewidth]{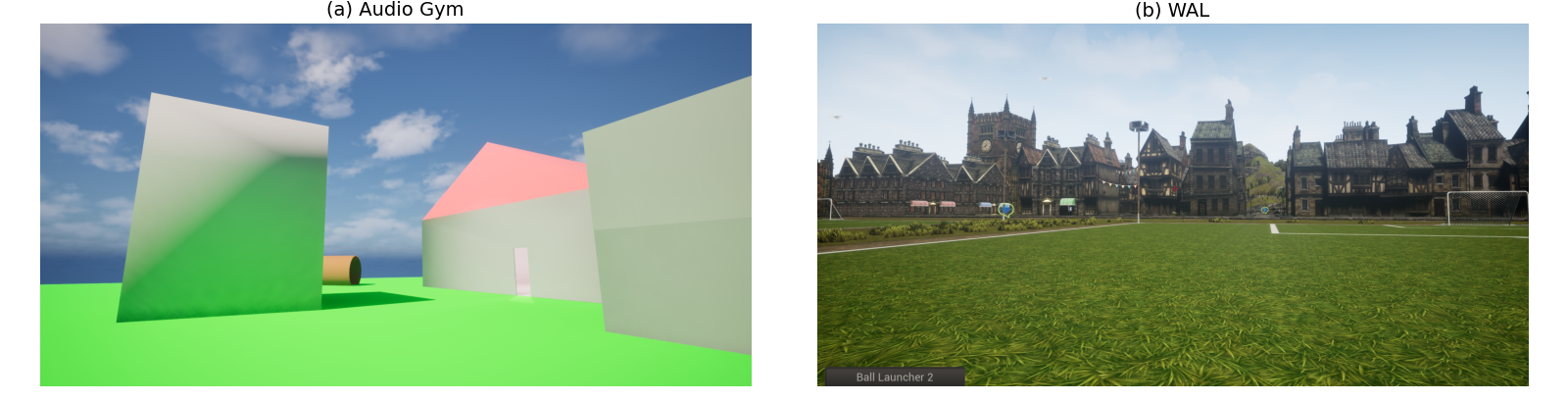}
  \caption{\textnormal{Image of the (a) Audio Gym and (b) WAL as seen by a player}}
  \label{fig:audio_gym_wal_image}
  \Description{}
\end{figure*}

We present our results for two maps: a custom map, which we call \emph{Audio Gym} (Fig. \ref{fig:audio_gym_wal_image}a) and the \emph{Wwise Audio Lab (WAL)} \cite{Audiokinetic2018} (Fig. \ref{fig:audio_gym_wal_image}b). The Audio Gym includes pathological features for sound propagation such as a 3D maze and an acoustically absorbing room coupled to a reflective room. The WAL is a typical map for video games with some buildings and large open spaces.

The Audio Gym spans $59 \times 8 \times 59\,\mathrm{m}$ and is represented on a $59 \times 8 \times 59$ grid at $1\,\mathrm{m}$ resolution. The WAL covers a larger area of $208 \times 10 \times 185\,\mathrm{m}$ and is discretized using a $1.2\,\mathrm{m}$ grid, resulting in a $173 \times 8 \times 154$ data grid.

\subsubsection{Models} \label{sec:results:setup:models}
We evaluate several model architectures.
\emph{RLF Euclidean}, \emph{RLF $G_{PSD}$}, \emph{RLF $G_{DIAG}$} respectively refer to the models in Eq.~\eqref{eq:euclidean_model}, \eqref{eq:full_riemannian}, and \eqref{eq:riemann_diag}. Those models can predict path distance and ER/LR levels.
\emph{MLP - Small} and \emph{MLP - Large} refer to two variants of Eq.~\eqref{eq:mlp_model}. \emph{MLP - Small} has two hidden layers with 32-32 units. \emph{MLP - Large} has three, with 128-64-32 units. Both feature ReLU activations and can predict path distance, levels and decay times. \emph{Dot-product} refers to the model in Eq.~\eqref{eq:dot_product_decay_times} and is used to predict ER/LR decay times only.

\subsubsection{Training and evaluation}  \label{sec:results:setup:training}
We generate several sets of distinct sources (see Sec.~\ref{sec:reciprocal_lf:practical:dataset}), divided into training/validation/test groups: 127/42/38 sources for the Audio Gym and 182/97/102 sources for the WAL. Wave simulations are carried out for each source position, forming our datasets. Although training, validation, and test sources are strictly non-overlapping, the adaptive sampling algorithm may introduce spatial correlations, as some regions can be sampled more densely than others. This may result in a slight underestimation of reconstruction error on the test set.

Each model is then trained for 20k epochs using the Adam optimizer. Learning rates were determined to ensure convergence during this interval. Latent embeddings are initialized as a positional embedding perturbed by small Gaussian noise: $\theta_{\text{init}}[i,j,k] = [i,j,k,0,\ldots,0] + 10^{-3}\varepsilon, \quad \varepsilon \sim \mathcal{N}(0,I)$. MLPs and linear layers use uniform initialization. All models use 32-bit precision floats. The \emph{validation set} is used to find which training iteration yields the best average error. We report the Mean Absolute Error (MAE) of the \emph{test set} at this iteration. Training times range from less than one hour (\emph{RLF Euclidean} and \emph{RLF $G_{DIAG}$}) to tens of hours (\emph{MLP - Large}) on two NVIDIA Quadro 8000 GPUs.

\subsection{Comparative study of model architectures}  \label{sec:results:comparative}

\begin{figure*}[htbp]
  \centering
  \includegraphics[width=\linewidth]{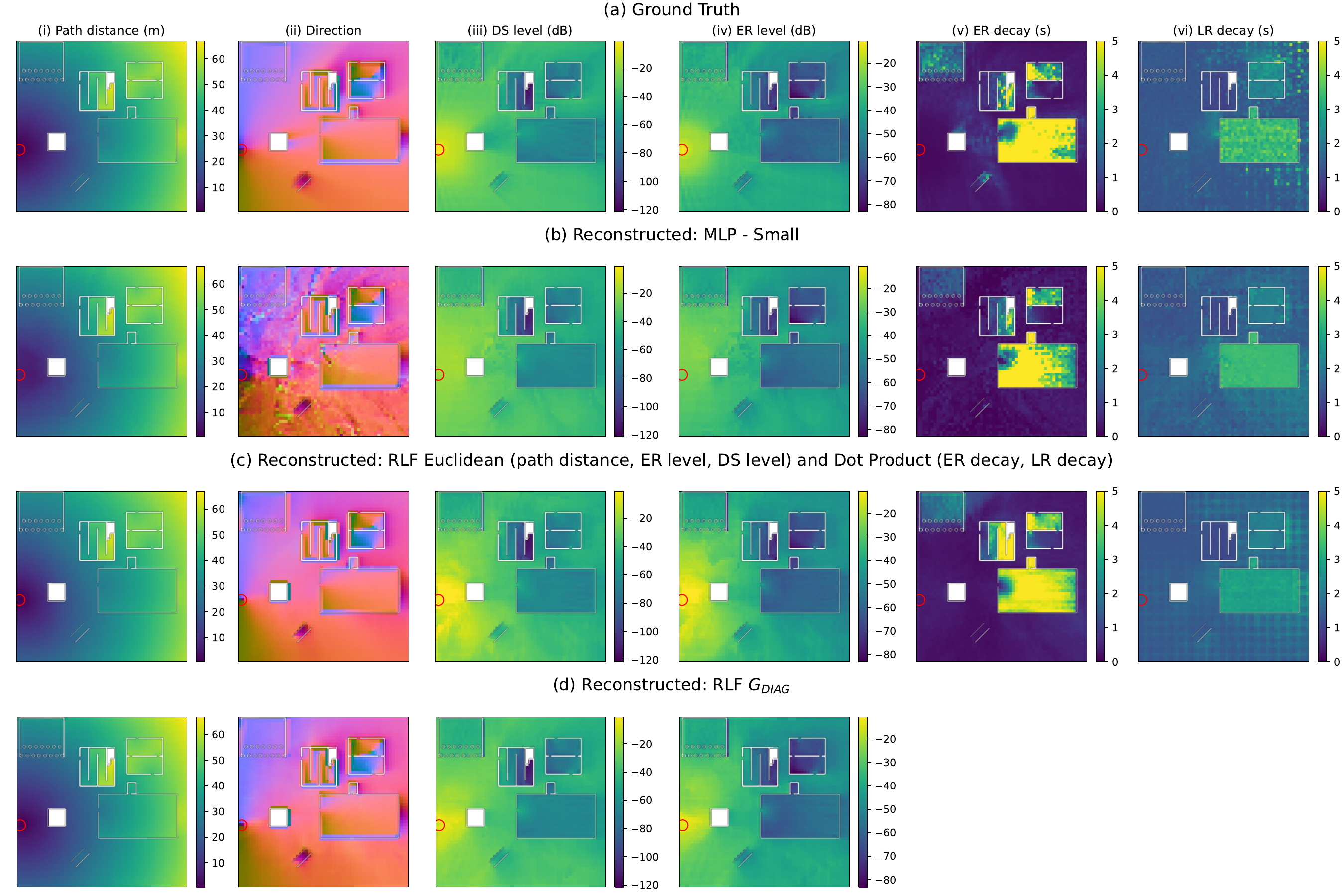}
  \caption{(a) Ground truth and (b - d) reconstructed fields. \textnormal{Fields are shown in the Audio Gym for a fixed source position (circle marker) that was \emph{not} seen during training, at a horizontal slice 1.3 m above ground. Thin white outlines indicate positions of walls. The DOA field in (ii) is visualized as an RGB image of the x-, y-, and z-components of the unit direction vector. Evaluated models are those described in Sec.~\ref{sec:results:setup:models}, with a latent space size fixed to $n=16$.}}
  \label{fig:ground_truth_and_reconstructed}
  \Description{}
\end{figure*}

\begin{figure*}[htbp]
  \centering
  \includegraphics[width=\linewidth]{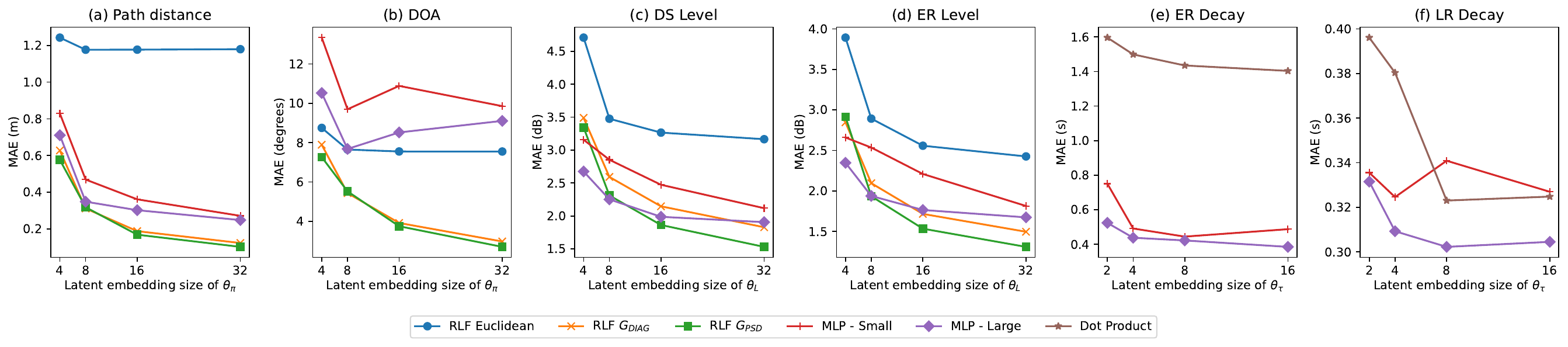}
  \caption{Mean Absolute Error (MAE) against latent embedding size \textnormal{for the Audio Gym map. See Sec.~\ref{sec:results:setup:models} for a description of the models.}}
  \label{fig:testmap_error_vs_lsize}
  \Description{}
\end{figure*}

Fig.~\ref{fig:ground_truth_and_reconstructed} shows examples of ground-truth and reconstructed fields for an unseen source position.
All models tend to reproduce the main structure of the acoustic fields well with some exceptions: MLP-based models exhibit noisy gradients leading to poor DOA estimation (Fig.~\ref{fig:ground_truth_and_reconstructed}(ii)), while the RLF Euclidean model shows localized noise in ER/LR levels near source positions.
Some high-frequency variations in DS and ER levels caused by shadowing or interference are smoothed by all models; however we find this has limited perceptual impact (see Sec.~\ref{sec:results:mushra}).
Finally, decay-time estimation, particularly for ER, is affected by noise in the ground truth itself, making these parameters harder to reconstruct. While this points to limitations in the simulation and signal processing pipeline (Sec.~\ref{sec:parametric_coding}), our empirical observations suggest that ER/LR decay times have a secondary perceptual impact compared to the other acoustic parameters.

Table~\ref{tab:summary_of_results} reports quantitative cost and performance metrics for both the \emph{Audio Gym} and \emph{WAL} scenes, and confirms the trends observed in the qualitative analysis.
Among the evaluated approaches, \emph{RLF $G_{\mathrm{PSD}}$} yields the most accurate estimates for path distance and sound levels.
\emph{RLF $G_{\mathrm{DIAG}}$} achieves comparable errors at a substantially lower inference cost.
MLP-based models perform well for level and decay-time estimation, but incur higher inference costs than \emph{RLF $G_{\mathrm{DIAG}}$}.
MLPs and dot-product models achieve similar errors for LR decay times, while the dot-product model is less effective at reproducing the noisy ER decay time field.
Errors in $L_{DS}$ and $L_{ER}$ can partially be attributed to regions where the levels are below -60 dB. Since game audio pipelines typically silence signals at this threshold for efficiency, these discrepancies are imperceptible.

\begin{figure*}[htbp]
  \centering
  \includegraphics[width=0.667\linewidth]{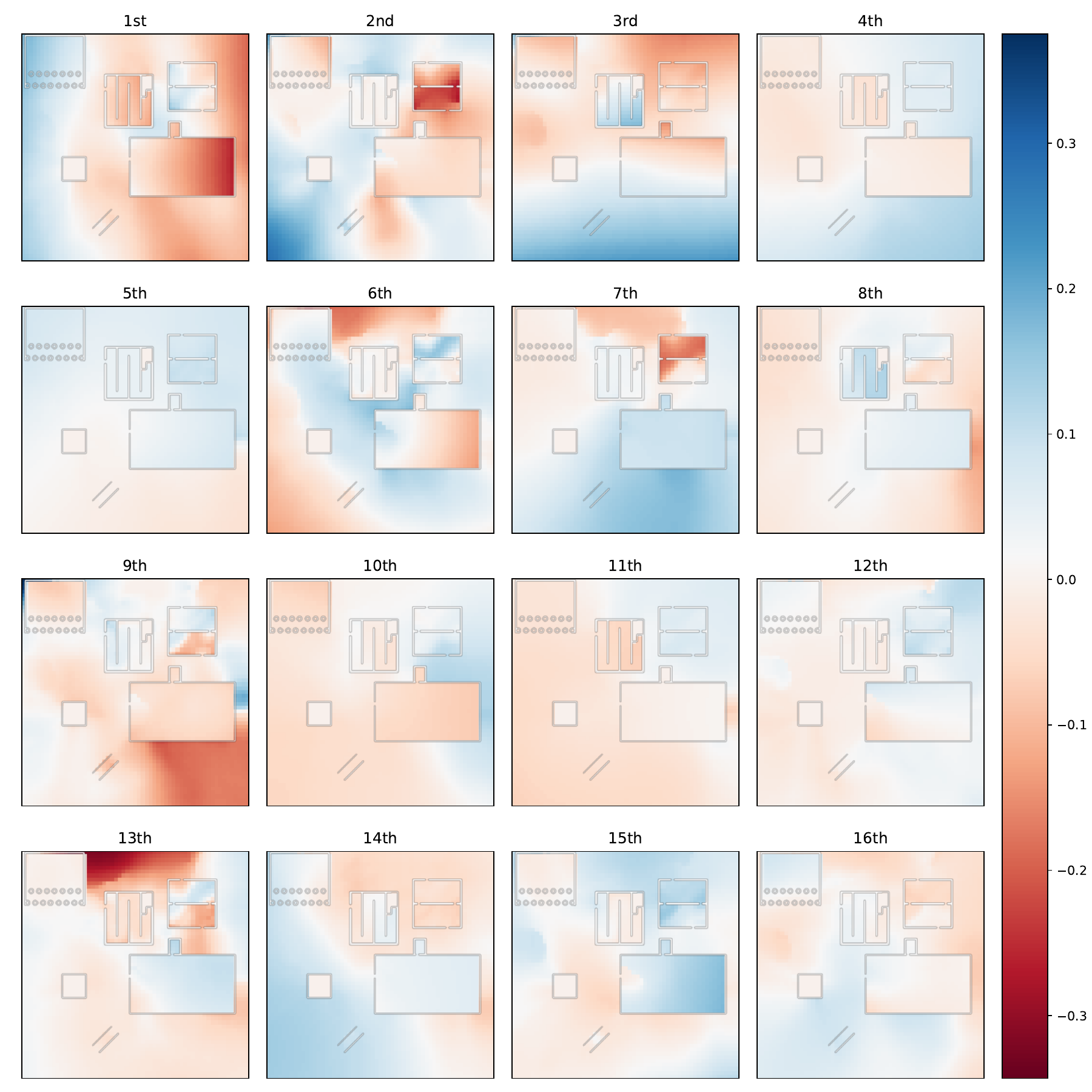}
  \caption{Individual components of latent embeddings $\theta_{\pi}$.\textnormal{This latent space is used to predict path distance for the Audio Gym map. The model is a $G_{DIAG}$ Riemannian RLF with $n=16$ described in Sec.~\ref{sec:results:setup:models}. Panels 1 - 16 depict the components of $\theta_{\pi}$ at a horizontal slice 1.3 m above the ground. Thin black outlines indicate positions of walls.}}
  \label{fig:latent_field}
  \Description{}
\end{figure*}

Finally, all proposed models drastically reduce raw memory requirements compared to wave coding methods. Although compression is beyond the scope of this paper, we note that wave coding can achieve substantial memory savings through delta-based compression that exploits spatial smoothness~\cite{Raghuvanshi2014}. Our latent fields exhibit similar spatial coherence, suggesting that comparable compression strategies could be applied (Fig.~\ref{fig:latent_field}).

\begin{table*}[b]
  \caption{Summary of model results and performance \textnormal{reported for both maps as [\emph{Audio Gym} / \emph{WAL}]. See Sec.~\ref{sec:parametric_coding:parameter_estimation} for the details of acoustic parameters. All latent embedding dimensions are set to $n=16$. Errors are not reported for the Wave Coding method as it is the ground truth used as target for other methods. \emph{Params} is the number of trainable parameters in the decoder. \emph{Memory} is the uncompressed data size needed to reconstruct one acoustic parameter family (see \ref{sec:reciprocal_lf:practical:shared_latent}), considering all values are coded as float32. \emph{FLOPs} is an approximation of the number of float operations needed for a single inference call. \emph{Latency} is the CPU inference time for 100 source-receiver pairs, which is a typical runtime query for games. We report the median inference time over 100 inferences. All measurements are performed with PyTorch on an Intel Xeon W-2255 CPU using a single thread. Reported FLOPs and latency are given for a single acoustic parameter and do not include costs associated to data access and trilinear interpolation, as those are required for all methods in the table.}}
  \label{tab:summary_of_results}
  \centering
  \small
  \setlength{\tabcolsep}{0pt} 
  \renewcommand{\arraystretch}{1.0}

  \begin{tabular*}{\textwidth}{@{\extracolsep{\fill}}l cccccc cccc @{}}
    \toprule
    \textbf{Model} &
    \multicolumn{6}{c}{\textbf{Mean Absolute Errors (MAE)}} &
    \multicolumn{4}{c}{\textbf{Costs}} \\
    \cmidrule(lr){2-7}\cmidrule(lr){8-11}
    &
    \textbf{$\pi$ (m)} &
    \textbf{DOA ($^\circ$)} &
    \textbf{$L_{DS}$ (dB)} &
    \textbf{$L_{ER}$ (dB)} &
    \textbf{$\tau_{ER}$ (s)} &
    \textbf{$\tau_{LR}$ (s)} &
    \textbf{Params} &
    \textbf{Memory} &
    \textbf{FLOPs} &
    \textbf{Lat. ($\mu$s)} \\
    \midrule
    
    Wave Coding  &
    -- &
    -- &
    -- &
    -- &
    -- &
    -- &
    0 &
    3.1 / 182 GB &
    0 &
    0\\
    
    RLF Euclidean  &
    1.2 / 0.94 &
    7.5 / 4.3 &
    3.3 / 5.1 &
    2.6 / 5.7 &
    -- &
    -- &
    0 &
    1.8 / 14 MB &
    47 &
    10\\
    
    RLF $G_{\mathrm{PSD}}$  &
    \textbf{0.17} / \textbf{0.34} &
    \textbf{3.8} / \textbf{3.2} &
    \textbf{1.9} / \textbf{4.2} &
    \textbf{1.5} / \textbf{2.6} &
    -- &
    -- &
    4.1k &
    1.8 / 14 MB &
    8.5k &
    93\\
    
    RLF $G_{\mathrm{DIAG}}$ &
    0.19 / 0.40 &
    3.9 / 3.3 &
    2.2 / 4.3 &
    1.7 / 2.8 &
    -- &
    -- &
    256 &
    1.8 / 14 MB &
    335 &
    46\\
    
    MLP - Small &
    0.36 / 0.42 &
    11 / 6.2 &
    2.5 / 4.3 &
    2.2 / 3.0 &
    0.49 / 0.76 &
    0.33 / 0.23 &
    2.1k &
    1.8 / 14 MB &
    8.3k &
    139\\
    
    MLP - Large &
    0.30 / 0.51 &
    8.5 / 7.8 &
    2.0 / 4.2 &
    1.8 / 3.0 &
    \textbf{0.39} / \textbf{0.73} &
    \textbf{0.30} / 0.22 &
    14.6k &
    1.8 / 14 MB &
    57.0k &
    247\\
    
    Dot Product &
    -- &
    -- &
    -- &
    -- &
    1.4 / 0.83 &
    0.32 / \textbf{0.19} & 
    0 &
    1.8 / 14 MB &
    32 &
    20\\
    
    \bottomrule
  \end{tabular*}
\end{table*}

\subsection{Influence of latent space size}  \label{sec:results:latent_space_ablation}
\label{sect:results_error_vs_lsize}
As expected, increasing the latent embedding dimension generally reduces reconstruction error (Fig.~\ref{fig:testmap_error_vs_lsize}).  
Riemannian RLFs benefit more from larger latent sizes and do not exhibit the early saturation observed in the Euclidean RLF.  
The \emph{MLP-Large} model achieves lower errors at small latent dimensions, but this trend reverses around $n=8$--$16$, where Riemannian models become more accurate.  
For both MLP models, improved field reconstruction does not translate to better gradient estimates (see Fig.~\ref{fig:testmap_error_vs_lsize}(b)).

\subsection{Subjective test} \label{sec:results:mushra}

Table~\ref{tab:summary_of_results} and Fig.~\ref{fig:ground_truth_and_reconstructed} indicate that small reconstruction errors remain in the proposed approach.
To evaluate whether these differences result in perceptible degradation compared to ground-truth simulation data, we conducted a MUSHRA-like (Multiple Stimuli with Hidden Reference and Anchor) study comparing the following systems:

\begin{itemize}[leftmargin=1.0em, style=nextline, noitemsep]
    \item \emph{Proposed method (\emph{RLF}).}
    Based on the results of Sec.~\ref{sec:results:comparative}, we adopt a 16-dimensional \emph{RLF $G_{\mathrm{DIAG}}$} model for path distance and level estimation, combined with an 8-dimensional dot-product model for decay times, as this configuration offers a good balance between accuracy and computational cost.
    
    \item \emph{Ground truth (\emph{GT}).}
    This corresponds to the wave coding method described in Sec.~\ref{sec:parametric_coding:rendering}.
    
    \item \emph{Low anchor (\emph{LA}).}
    A simplistic system based on free-space propagation, independent of scene geometry. Distance attenuation slopes and offsets for direct sound and early reflections are obtained from a linear fit to ground-truth data, to avoid overly salient level differences relative to the other systems.
\end{itemize}
Other real-time methods from Sec.~\ref{sec:related_work:simulation} were not included as these approaches heavily rely on subjective authoring, making the result dependent on the designer and not reproducible.

The RLF, GT, and LA systems are implemented in an interactive runtime environment for the \emph{Audio Gym} map, using the Godot game engine~\cite{Godot2024} and the Wwise audio middleware~\cite{Audiokinetic2025}. Both the RLF and GT methods are rendered using the method in Sec.~\ref{sec:parametric_coding:rendering} from their respective acoustic parameter fields.

For each system, we record synchronized audio and video using a fixed source position and receiver trajectory across five acoustic scenes spanning a wide range of conditions: a large reverberant building (Scene 1), an outdoor environment (Scene 2), a maze (Scene 3), a forest of cylindrical reflectors (Scene 4), and a coupled-room configuration connecting a highly absorptive space to a highly reflective one (Scene 5). Sound stimuli were gunshots and speech.
Twenty-eight game audio professionals rated the realism of the three systems on a 0–100 scale, based on the pre-recorded videos.

Ratings were analyzed using a linear mixed-effects model with system, scene, and their interaction as fixed effects, and participant-specific random intercepts and slopes. Mean scores are reported in Table \ref{tab:mushra_results}.
Pairwise comparisons reveal that both the proposed RLF model and the ground-truth wave coding system are rated significantly higher than the low anchor (both $p < 0.001$, large effect sizes with Cohen’s $d \approx 1.4$ \cite{Cohen1988}). In contrast, no statistically significant difference is observed between RLF and ground truth ($p = 0.41$, negligible effect size), indicating systems are perceived as equally realistic. This trend is consistent in all the acoustic scenes tested.

\begin{table}[h!]
\centering
\caption{Mean scores by system and scene, and corresponding 95\% confidence intervals}
\label{tab:mushra_results}
\small
\begin{tabular}{
  l
  S[table-format=2.1] @{${}\pm{}$} S[table-format=2.1]
  S[table-format=2.1] @{${}\pm{}$} S[table-format=2.1]
  S[table-format=2.1] @{${}\pm{}$} S[table-format=2.1]
}
\hline
 & \multicolumn{2}{c}{\textbf{RLF}}
 & \multicolumn{2}{c}{\textbf{GT}}
 & \multicolumn{2}{c}{\textbf{LA}} \\
\hline
Overall
 & 61.8 & 3.8
 & 62.6 & 3.9
 & 27.7 & 4.1 \\
\hline
Scene 1
 & 60.3 & 9.8
 & 56.1 & 9.5
 & 20.7 & 7.3 \\
Scene 2
 & 54.6 & 10.0
 & 60.1 & 10.4
 & 21.1 & 8.6 \\
Scene 3
 & 72.5 & 5.3
 & 69.0 & 7.1
 & 15.1 & 6.5 \\
Scene 4
 & 56.4 & 9.8
 & 62.1 & 9.4
 & 43.5 & 10.4 \\
Scene 5
 & 65.4 & 7.4
 & 65.6 & 8.6
 & 38.4 & 9.8 \\
\hline
\end{tabular}
\end{table}

\section{Conclusion} \label{sec:conclusion}
In this paper, we introduced Reciprocal Latent Fields (RLF), a framework for modeling sound propagation using latent spatial embeddings and physically motivated decoders. Our evaluation across architectures and latent dimensionalities shows that Riemannian RLFs can accurately predict sound propagation for unseen source–receiver configurations. This accuracy is achieved from a limited number of precomputed wave simulations and with modest latent dimensionality. The approach yields orders-of-magnitude reductions in raw memory usage compared to wave coding methods, while maintaining low inference costs compatible with real-time game applications.

Our work has two main limitations. First, we did not implement spatial compression on the latent fields, although a comparison based on compressed memory footprints would be more representative of practical applications. More fundamentally, the current RLF framework is restricted to static geometry, as embeddings need to be trained on simulations for a specific map. Although there are limited workarounds such as supporting a small number of predefined portals \cite{Raghuvanshi2021}, fully dynamic environments are not addressed.

Our primary direction for future work is to overcome this limitation by extending RLFs to dynamic geometry, allowing the latent representations to adapt to scene changes at runtime. Beyond acoustics, RLFs also provide a general framework for encoding reciprocal quantities that are influenced by geometry. For example, the path-distance field and its gradients could support applications such as pathfinding or efficient line-of-sight queries by comparing it with the direct Euclidean distance.

\begin{acks}
We thank Marc-André Carbonneau and Jean-François Guay for their vision and foundational work on this project. We are grateful to our colleagues at Ubisoft, Audiokinetic, and Sony Interactive Entertainment for performing the subjective listening tests.
\end{acks}


\bibliographystyle{ACM-Reference-Format}
\bibliography{sample-base}


\end{document}